# MindGrab for BrainChop: Fast and Accurate Skull Stripping for Command Line and Browser

Armina Fani[1], Mike Doan[1], Isabelle Le[1], Alex Fedorov[2], Malte Hoffmann[3], Chris Rorden[4], Sergey Plis[1]

[1] Tri-Institutional Center for Translational Research in Neuroimaging and Data Science (TReNDS), Georgia State University, Georgia Institute of Technology, Emory University, Atlanta, GA, USA; [2] Emory University, Atlanta, GA, USA; [3] Harvard University, Cambridge, MA, USA; [4] University of South Carolina, Columbia, SC, USA

**Abstract**
Deployment complexity and specialized hardware requirements hinder the adoption of deep learning models in neuroimaging. We present MindGrab, a lightweight, fully convolutional model for volumetric skull stripping across all imaging modalities. MindGrab's architecture is designed from first principles using a spectral interpretation of dilated convolutions, and demonstrates state-of-the-art performance (mean Dice score across datasets and modalities: 95.9 ± 1.6), with up to 40-fold speedups and substantially lower memory demands compared to established methods. Its minimal footprint allows for fast, full-volume processing in resource-constrained environments, including direct in-browser execution. MindGrab is delivered via the BrainChop platform as both a simple command-line tool (pip install brainchop) and a zero-installation web application (brainchop.org). By removing traditional deployment barriers without sacrificing accuracy, MindGrab makes state-of-the-art neuroimaging analysis broadly accessible.

## 1. Introduction

The transformative potential of deep learning in neuroimaging is increasingly hampered by a critical, yet often overlooked, barrier: deployment complexity. While novel architectures demonstrate state-of-the-art performance, their practical adoption by clinicians and researchers is severely limited by immense technical requirements (Renton et al., 2024; Gronenschild et al., 2012). These models often demand specialized hardware like high-end NVIDIA GPUs, convoluted software installations, and command-line expertise, effectively excluding the vast majority of their intended users. This disparity creates a paradox where our most powerful analytical tools remain inaccessible for routine point-of-care or research applications.

The specialized hardware required for deployment of these models is often only available through cloud-based platforms, which introduces a separate, frequently insurmountable, obstacle: data privacy. Strict institutional policies and regulations, such as HIPAA, prohibit the transfer of protected health information (PHI) to third-party servers, rendering most cloud solutions unsuitable for clinical data (Plis et al., 2016; Kaissis et al., 2020). This leaves the field at an impasse: local deployment is too complex, and cloud deployment is too insecure for many institutions under current regulatory constraints, hindering progress and compromising the quality of downstream analyses.

This challenge is particularly acute for foundational preprocessing tasks like skull stripping– the removal of non-brain tissue from neuroimaging scans. For decades, the field has relied on classical methods like BET (Smith, 2002) and ROBEX (Iglesias et al., 2011), which, despite their utility, often struggle with variations in image contrast and quality. Deep learning models, typically based on large U-Net architectures (Ronneberger et al., 2015), offered a leap in accuracy. More recently, the challenge of generalizing across different scanners and modalities was elegantly solved by training models entirely on synthetic data (Billot et al., 2023; Gopinath et al., 2024). The state-of-the-art method, SynthStrip (Hoopes et al., 2022), leverages this strategy to achieve unprecedented robustness. However, its large, parameter-heavy architecture imposes many practical challenges. Its deployment requires navigating a fragile ecosystem of software dependencies– a task demanding a level of systems administration expertise that is orthogonal to clinical and scientific practice. Furthermore, its computational and memory footprint, while manageable on a dedicated workstation, renders the architecture unsuitable for deployment in resource-constrained environments such as web browsers and handheld devices, which are becoming ubiquitous in clinical and research settings. These hurdles create significant friction for adoption, effectively limiting the immediate use of such powerful tools by clinicians at the point of care and researchers focused on analysis, not software installation.

To resolve this tension between accuracy and accessibility, we present MindGrab, an extremely efficient deep learning model for skull stripping. MindGrab achieves Dice scores that are highly comparable to SynthStrip by favoring higher precision (fewer false positives) in its segmentation strategy. This accuracy is achieved with a model that has 95% fewer parameters, an efficiency that translates into dramatic real-world performance gains, including up to 40-fold speedups and substantially lower memory usage on both high-end GPUs and consumer-grade hardware, enabling novel deployment modalities such as in-browser execution via our BrainChop platform (Plis et al., 2024). This architecture was designed from first principles guided by a spectral analysis of dilated convolutions. We demonstrate that this design, which systematically reduces spatial frequencies, is key to MindGrab's success by validating its performance against alternative dilation patterns. MindGrab thus provides the neuroimaging community with a tool that delivers state-of-the-art performance and is immediately usable without the traditional barriers of software installation or hardware dependency.

## 2. Methods

### 2.1 Model Architecture and Dilation Schedule Design

The design of dilated convolution architectures, such as MeshNet by Fedorov et al. (2017), has proven highly effective for patch-based analysis, where a receptive field of $69^3$ voxels is well-suited for processing sub-volumes. This strategy is suboptimal for single-pass, whole-volume analysis, as it risks stitching artifacts and cannot leverage global context. Simply extending traditional dilation schedules (Yu and Koltun, 2016) to cover a full volume (e.g., $256^3$) is inefficient and offers no clear guidance for creating a memory-light model. This motivated our move from a purely spatial receptive field analysis to a spectral perspective.

This spectral view provides a principled design path. In the frequency domain, dilation acts as a tuning mechanism. Introducing gaps between kernel weights creates a periodic filter that replicates its Fourier response across k-space (see Figure 1). This allows a single, small kernel to become sensitive to multiple, widely separated frequency bands without increasing its parameter count. Increasing the dilation rate creates a denser tiling of these sensitivity bands, broadening the range of spatial frequencies the layer can interact with. In essence, the dilation schedule determines which frequency bands the network can "see," while the learned weights determine the response within those bands. This principle is the foundation of our architectural design.

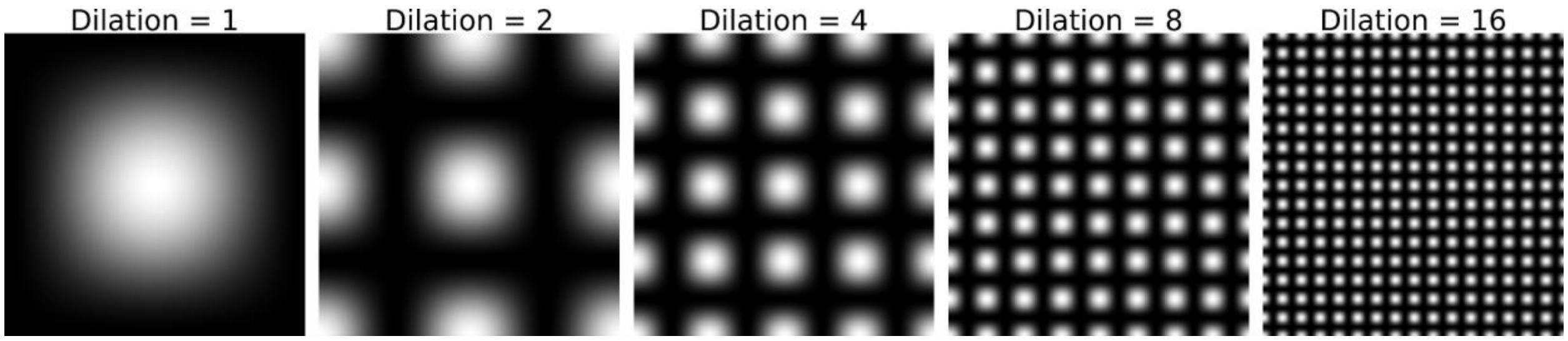

**Figure 1. k-Space Magnitude Envelopes of a 3x3 Kernel with Different Dilations.** Increasing dilation produces progressively denser spectral replication, broadening the range of spatial frequencies the filter can interact with while keeping the kernel size unchanged.

To exploit this property, we group layers (each a sequence of convolution, normalization, and Gaussian Error Linear Unit (GeLU) activation function (Hendrycks and Gimpel, 2023) into 5-layer dilation blocks that act as controlled spectral bottlenecks. Two complementary schedules are considered:

- Increasing: 1→2→4→8→16 (denoted ◀) — progressively broadens frequency support (sharpening).
- Decreasing: 16→8→4→2→1 (denoted ▶) — progressively contracts frequency support (blurring).

Because all layers are isometric (have the same spatial dimensions) and use the same channel count, the only variable that changes is the dilation pattern itself. This allows us to attribute performance differences directly to spectral structure rather than depth, capacity, or skip connections. This ordering of dilations produces a multiscale flow similar in spirit to a Laplacian pyramid, achieved without changing image resolution.

MindGrab is built from five consecutive decreasing blocks (▶▶▶▶▶), each containing five 3×3×3 convolutions with 15 channels, followed by a final 1×1×1 projection layer (26 layers total). The network uses parameter-free per-volume z-scoring and no biases. During inference, only one activation map is stored at a time, enabling deployment on memory-limited hardware and direct in-browser execution as showcased in our BrainChop platform.

## 2.2 Synthetic Data Generation for Training

MindGrab was trained *exclusively on synthetic data* generated with Wirehead (Doan and Plis, 2025), a data generation pipeline leveraging SynthSeg (Billot et al., 2023) to continuously produce diverse synthetic (brain image, label) pairs. SynthSeg begins with an anatomical label map and applies a series of randomized augmentations, including spatial deformations, intensity variations, resolution randomization, and simulation of imaging artifacts such as blurring and noise. This process enables the model to learn domain-agnostic features, making it highly robust to variations in real-world scans. Our input label set to Wirehead included 171 volumes, each with 39 standard anatomical FreeSurfer and non-brain labels. This included 131 label maps from the publicly available SynthStrip dataset (Hoopes et al., 2022) plus 40 cropped variants generated by truncating superior and inferior brain areas to simulate scans with sharp cutoffs.

All synthetic data were preprocessed with 2nd-98th percentile quantile normalization. Brain-specific labels were merged into a binary mask with smoothed edge boundaries. MindGrab was trained on approximately 250k synthetic samples using the Adam optimizer, soft-dice loss, 50 cycles of OneCycle-LR, and a batch size of one.

### 2.3 Evaluation Datasets

The validation dataset was derived from the multimodal benchmark compiled for SynthStrip evaluation, comprising 606 adult images from eight public datasets (Hoopes et al., 2022; Biomedical Image Analysis Group; Greve et al., 2021). The IXI dataset contributed 50 T1-weighted (T1w), 50 T2-weighted (T2w), 50 proton density-weighted (PDw), 50 magnetic resonance angiography (MRA), and 32 diffusion-weighted imaging (DWI) scans. The FSM subset included 38 T1w, 36 T2w, 32 PDw, and 32 quantitative T1 maps (qT1). SynthStrip provided 43 pseudo-continuous ASL (PCASL) T1w scans and 43 2D echo planar imaging (EPI) acquisitions (Harms et al., 2018; Juttukonda et al., 2021). Clinical stacks of thick image slices from the QIN glioblastoma dataset contributed 54 T1w, 39 T2w, and 17 T2-FLAIR volumes (Mamonov and Kalpathy-Cramer, 2016; Prah et al., 2015; Clark et al., 2013). The CERMEP-IDB-MRXFDG (CIM) database provided 20 brain CT and 20 PET scans (Merida et al., 2021). All samples were accompanied by a silver standard (synthetically generated as an average of multiple automatically produced masks from different methods, but not human labels) ground-truth brain mask, and all data were resampled to $256^3$ shape with 1-mm isotropic resolution.

## 3. Results

The following subsections evaluate the performance of MindGrab against baseline models across multiple imaging modalities in terms of accuracy, robustness, and computational efficiency. We also evaluate MindGrab⚡, which is the same model as MindGrab, but applied to cropped versions of the input, where the empty space around the head is discarded before application. We further isolate the contribution of dilation order and normalization strategy to quantify their roles in the model's performance.

### 3.1 Comparison with Existing Models

We evaluate the similarity between computed and ground-truth brain masks for MindGrab, SynthStrip, ROBEX, BET, and MindGrab⚡ by reporting average Dice, precision, and recall scores. Where given, the significance of MindGrab scores is determined by two-sided Wilcoxon signed-rank significance tests.

**Table 1**: Dice Comparison of Skull Stripping Models ↑

| Modalities | MindGrab⚡ | MindGrab | SynthStrip | ROBEX | BET |
|---|---|---|---|---|---|
| FSM T1w | 97.6 ± 0.3 ◦ | 97.5 ± 0.3 | 97.8 ± 0.3 ◦ | 96.0 ± 0.7 • | 66.8 ± 8.9 • |
| IXI T1w | 97.3 ± 0.4 • | 97.5 ± 0.4 | 97.1 ± 0.5 • | 96.1 ± 0.8 • | 88.2 ± 6.9 • |
| FSM qT1 | 97.5 ± 0.4 ◦ | 97.4 ± 0.4 | 97.7 ± 0.2 ◦ | 81.7 ± 11.8 • | 68.2 ± 4.0 • |
| ASL T1w | 96.8 ± 0.6 • | 97.3 ± 0.5 | 97.3 ± 0.5 · | 96.8 ± 1.1 • | 89.4 ± 5.7 • |
| FSM T2w | 97.3 ± 0.5 ◦ | 97.1 ± 0.5 | 97.8 ± 0.3 ◦ | 93.0 ± 1.8 • | 92.0 ± 5.7 • |
| IXI T2w | 97.1 ± 0.7 ◦ | 97.0 ± 0.7 | 96.6 ± 0.5 • | 91.4 ± 2.6 • | 92.5 ± 3.5 • |
| IXI PDw | 97.2 ± 0.7 ◦ | 96.9 ± 0.6 | 96.7 ± 0.5 • | 94.5 ± 1.3 • | 94.7 ± 1.9 • |
| IXI MRA | 94.1 ± 1.3 • | 96.7 ± 0.7 | 97.4 ± 0.5 ◦ | 73.9 ± 8.4 • | 86.7 ± 9.5 • |
| FSM PDw | 96.9 ± 0.5 ◦ | 96.4 ± 0.5 | 97.6 ± 0.3 ◦ | 95.6 ± 1.0 • | 87.4 ± 6.9 • |
| QIN FLAIR | 96.1 ± 0.4 · | 96.0 ± 0.5 | 96.0 ± 0.5 · | 93.0 ± 4.1 • | 95.4 ± 1.2 • |
| CIM CT | 95.7 ± 1.5 · | 95.9 ± 1.2 | 95.3 ± 1.0 • | 73.8 ± 2.4 • | 41.5 ± 4.4 • |
| QIN T1w | 94.6 ± 2.0 · | 94.9 ± 1.5 | 95.8 ± 1.0 ◦ | 92.9 ± 3.6 • | 92.8 ± 3.1 • |
| CIM PET | 95.0 ± 1.3 ◦ | 94.7 ± 1.4 | 95.0 ± 1.0 · | 91.9 ± 3.5 • | 89.2 ± 5.9 • |
| IXI DWI | 94.0 ± 1.4 ◦ | 93.7 ± 1.7 | 95.6 ± 0.9 ◦ | 87.2 ± 6.3 • | 93.9 ± 1.3 · |
| QIN T2w | 93.4 ± 1.9 · | 93.5 ± 1.8 | 95.0 ± 1.1 ◦ | 87.1 ± 6.7 • | 89.3 ± 4.4 • |
| ASL EPI | 92.3 ± 1.1 · | 92.4 ± 0.9 | 95.2 ± 1.0 ◦ | 80.8 ± 6.4 • | 94.6 ± 1.1 ◦ |

• = MindGrab's statistical superiority ◦ = listed model's superiority · = no statistical difference

**Table 1. Skull Stripping Performance Comparison Across Multimodal Datasets**. Metrics show the average Dice score ± standard deviation between computed and ground-truth brain masks. Rows are ordered by descending MindGrab Dice score. MindGrab is applied to $256^3$ inputs. MindGrab and MindGrab⚡ achieve comparably competitive or superior performance, significantly outperforming ROBEX universally and BET in most cases, with mixed results against SynthStrip.

The Dice comparison in Table 1 demonstrates that MindGrab achieves average Dice scores exceeding ROBEX and BET and highly comparable to SynthStrip. MindGrab significantly outperforms ROBEX across all datasets and is significantly better than BET in all but two cases: IXI DWI (no significant difference) and ASL EPI (BET performs better). Compared to SynthStrip, MindGrab is significantly better in four categories, shows no significant difference in three, and is significantly worse in nine, remaining within 3% of SynthStrip's Dice score, indicating competitive overall performance.

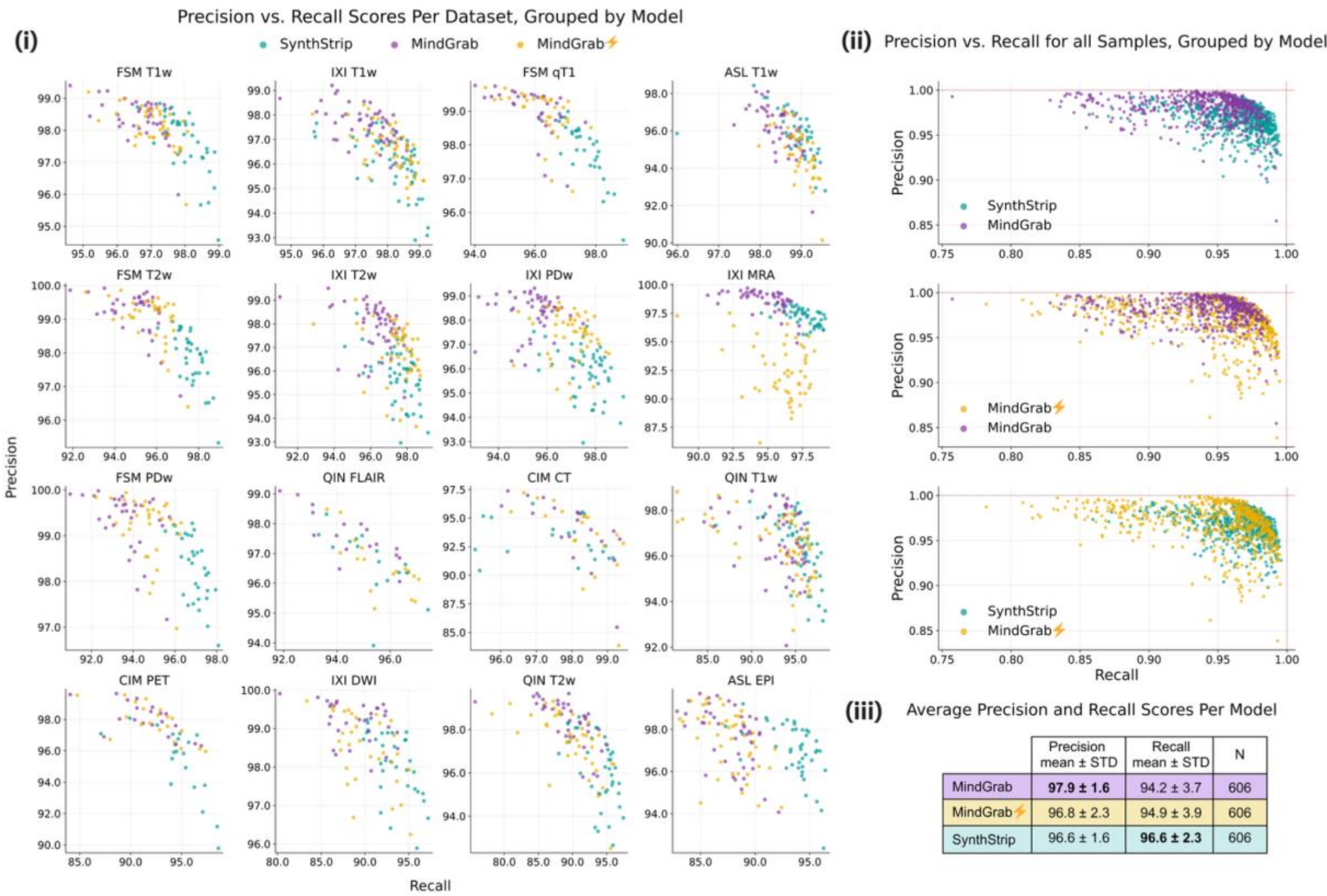

| | Precision mean ± STD | Recall mean ± STD | N |
|---|---|---|---|
| MindGrab | **97.9 ± 1.6** | 94.2 ± 3.7 | 606 |
| MindGrab⚡ | 96.8 ± 2.3 | 94.9 ± 3.9 | 606 |
| SynthStrip | 96.6 ± 1.6 | **96.6 ± 2.3** | 606 |

**Figure 2. Precision and Recall Comparison Across Evaluation Datasets**. Panel (i) presents a scatter plot grid of precision (y-axis) vs. recall (x-axis) scores for the 16 evaluated datasets. Each dot represents a single sample, with SynthStrip scores in blue, MindGrab in purple, and MindGrab⚡ in yellow. MindGrab consistently displays higher precision, while SynthStrip generally achieves higher recall. MindGrab⚡ occupies an intermediate position, offering a balance between the two. Panel (ii) provides a global view of the precision-recall relationship by grouping all samples by model. The top graph in this sub-figure compares SynthStrip and MindGrab, the middle compares MindGrab and MindGrab⚡, and the bottom compares SynthStrip with MindGrab⚡. Panel (iii) is a summary table that presents the mean precision and recall (± standard deviation) and the total sample count for the three models, confirming the trends observed in the scatter plots. The highest score for each metric is bolded.

Figure 2 gives insight into precision and recall trends for MindGrab, MindGrab⚡, and SynthStrip. MindGrab (purple) generally achieves higher precision scores, whereas SynthStrip (blue) demonstrates higher recall across the majority of datasets. Notable exceptions to this trend are observed in datasets CIM CT, ASL T1w, QIN T1w, and QIN FLAIR, where the performance relationship between the models is either more equivalent or less distinct. MindGrab⚡'s performance (yellow) typically falls between these two models, indicating a balance between the precision and recall trade-off.

This relationship is reinforced by the aggregate data presented in Panel (ii), which plots all samples by model. MindGrab trends towards higher precision, whereas SynthStrip favors higher recall (top subplot). MindGrab⚡ mostly follows MindGrab's performance, with slight variability in precision and marginally higher recall. This positions MindGrab⚡ at a slightly higher precision and lower recall than SynthStrip. The quantitative summary in Panel (iii) further supports these observations. MindGrab has the highest precision of the three, with a score of 97.9 ± 1.6 compared to SynthStrip's 96.6 ± 1.6 and MindGrab⚡'s 96.8 ± 2.3. SynthStrip has the highest average recall with a score of 96.6 ± 2.3, compared to MindGrab's 94.3 ± 3.7, and MindGrab⚡'s 94.9 ± 3.9. (See supplementary Tables S1 and S2 for Mean Surface Distance and 95th Percentile Hausdorff Distance scores).

### 3.2 Qualitative Evaluation of Brain-Mask Boundaries

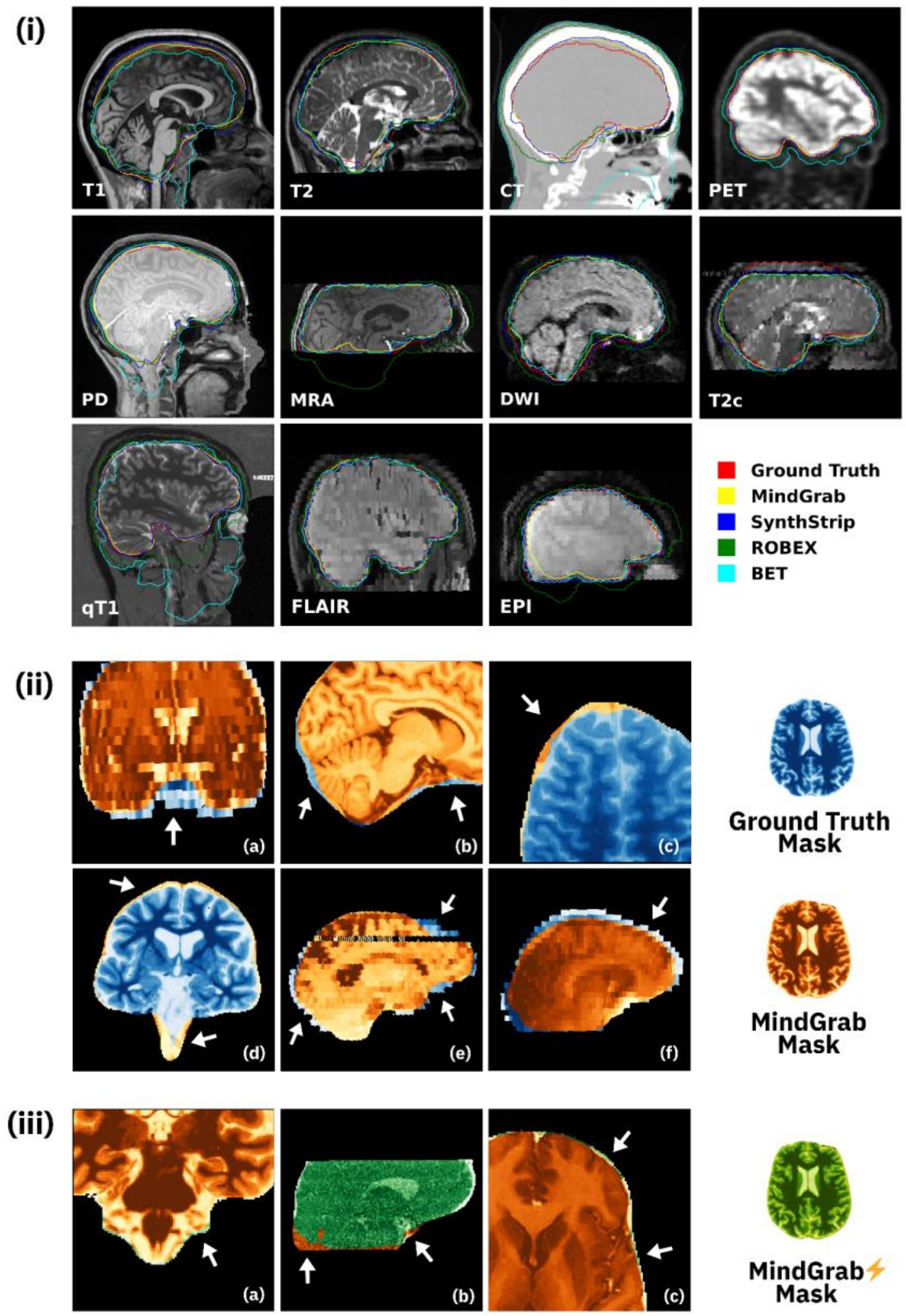

**Figure 3**. **Qualitative Comparison of Skull Stripping Results Across Different Methods and Imaging Modalities**. In Panel (i), segmentation boundaries are displayed as colored contours overlaid on representative sagittal slices from different imaging modalities: MindGrab (yellow), SynthStrip (blue), ROBEX (green), and BET (cyan). Ground truth contours are shown in red. Note the variable performance of classical methods (ROBEX, BET) and the generally comparable accuracy between MindGrab and SynthStrip. Panel (ii) details instances of the rare MindGrab error by highlighting differences in ground truth and MindGrab mask overlap. MindGrab masks are orange, and ground truth masks are blue. MindGrab shows cases of both over- and under-segmentation. Panel (iii) compares MindGrab and MindGrab⚡, with MindGrab⚡ masks shown in green. While MindGrab⚡ generally shows minimal differences from MindGrab, it tends to under-segment in regions with sharp intensity transitions.

Panel (i) of Figure 3 qualitatively compares MindGrab, SynthStrip, ROBEX, and BET through superimposed segmentation and the silver-standard ground-truth contours overlaid on input images. Classical methods like ROBEX (green) and BET (cyan) show variable performance, with notable instances of both over- and under-segmentation. SynthStrip (blue) demonstrates reliable boundary alignment but exhibits minor over-segmentation inferior to the medial prefrontal cortex. MindGrab (yellow) achieves comparable efficacy with reduced precision near high-contrast boundaries, as seen in the MRA example. Both SynthStrip and MindGrab fail to segment regions beyond abrupt intensity transitions, as seen superior to a dark artifact in the T2c image.

Panel (ii) details localized segmentation errors associated with MindGrab. Subpanels (a) and (f) show a conservative segmentation approach, characterized by a slight undersegmentation of the brain when it is in close proximity to the skull or dura. This behavior is also present in subpanel (b), where the arrows highlight a subtle undersegmentation of the cerebellum's inferior boundary. Conversely, subpanels (c) and (d) reveal cases of slight over-segmentation, where the predicted mask is marginally larger than the ground truth.

Panel (iii) compares the segmentation outputs of MindGrab and MindGrab⚡. As shown in subpanels (a) and (c), the typical difference between the two models is minimal and often visually imperceptible. However, on more challenging datasets, MindGrab⚡ can exhibit instances of greater under-segmentation, as seen in subpanel (b). These larger deviations from the ground truth are particularly notable in images with artifacts that have sharp boundary transitions. This behavior is consistent with the lower precision observed in the quantitative analysis section earlier.

### 3.3 Computational Efficiency Analysis

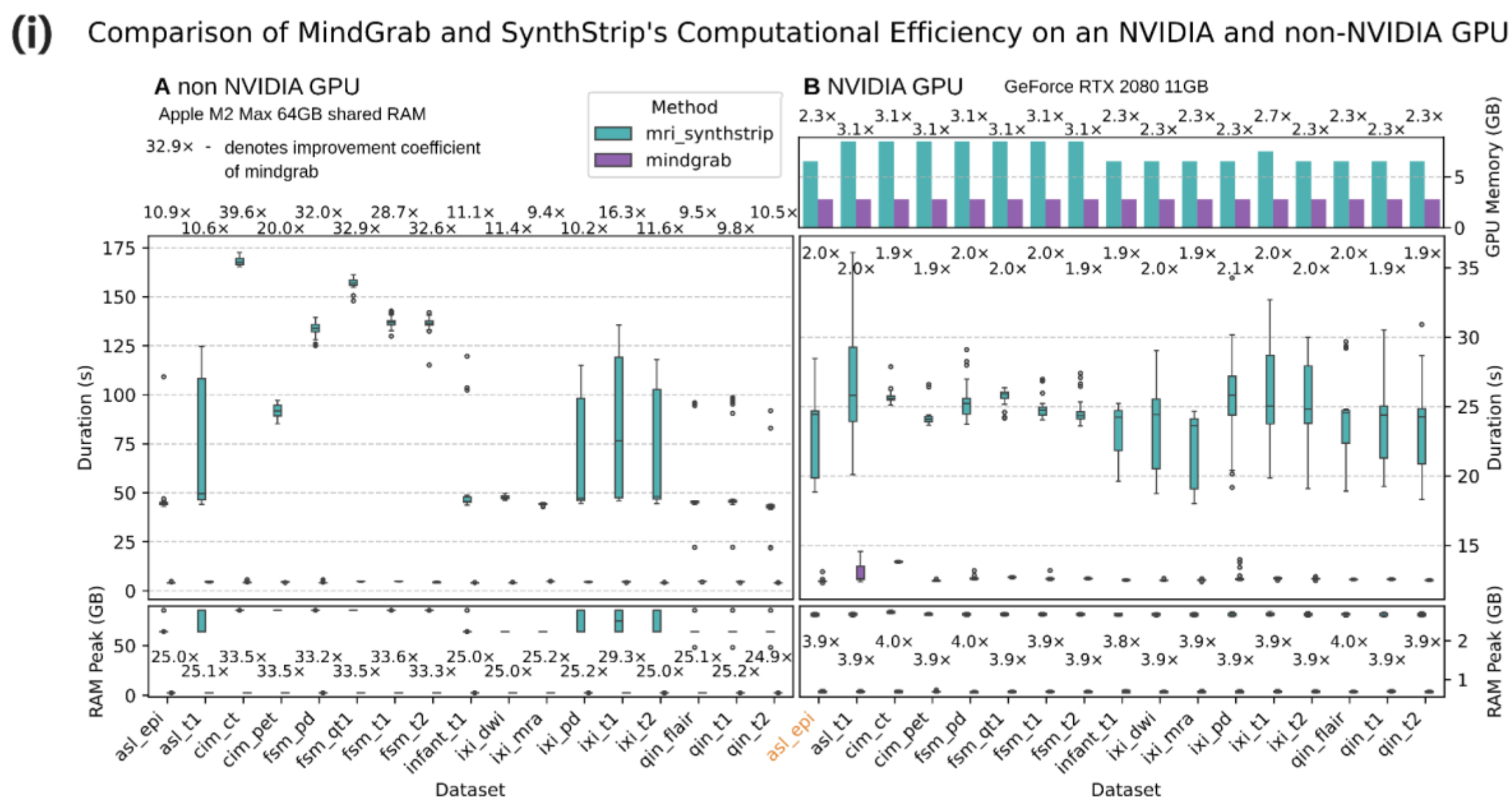

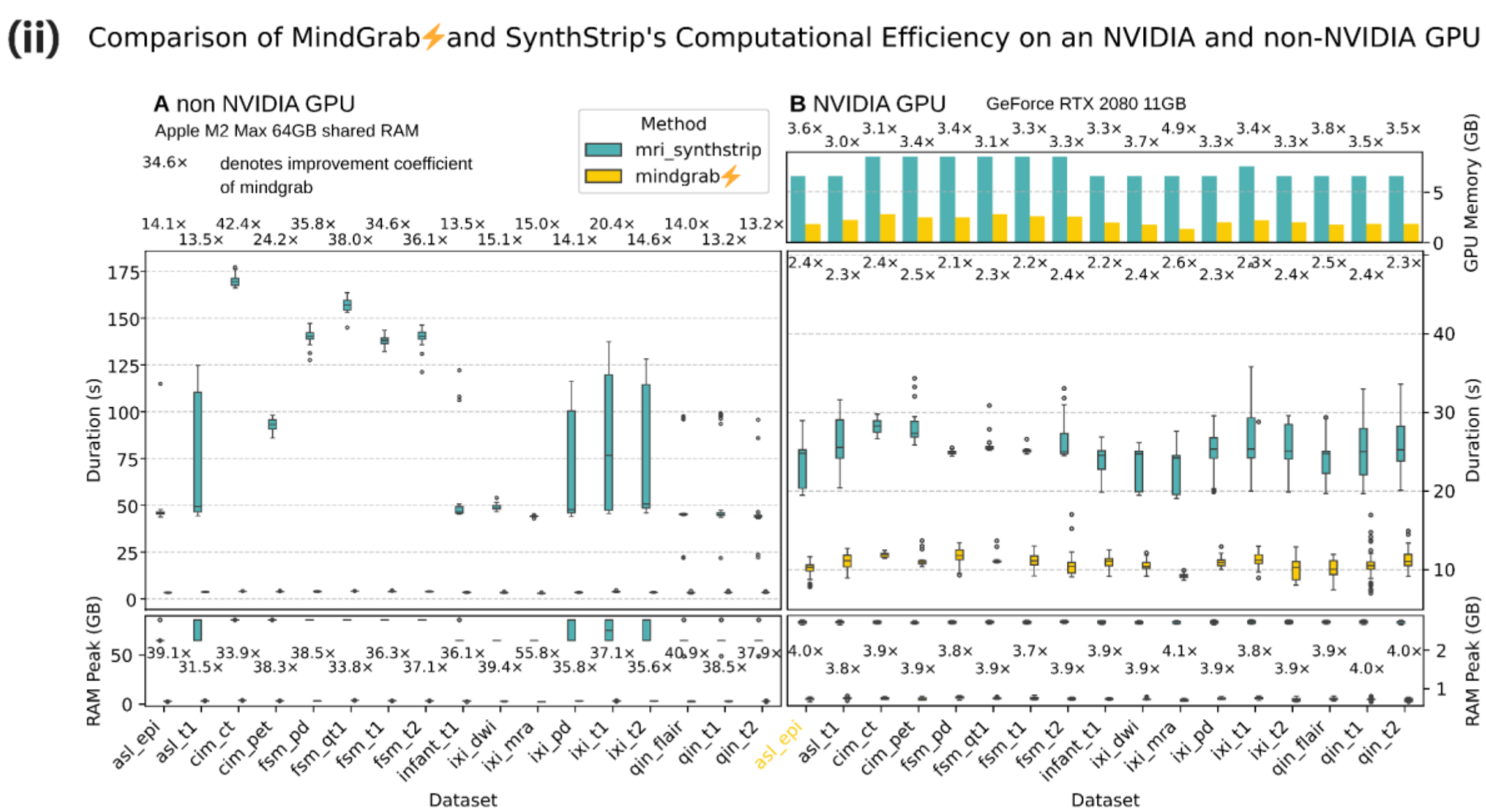

**Figure 4**. **A Comparative Analysis of MindGrab, MindGrab⚡ and SynthStrip's Computational Efficiency for Processing the 16 Evaluation Datasets.** Panel (i) compares MindGrab and SynthStrip, and Panel (ii) compares MindGrab⚡ and SynthStrip. Performance is benchmarked for an Apple M2 Max (64GB shared RAM, A) and an NVIDIA GeForce RTX 2080 (11GB, B). The figure compares peak RAM usage (GB, bottom subplot) and run duration (s, middle subplot), with GPU memory usage (GB, top subplot) also shown for the NVIDIA GPU (as GPU memory is shared on Apple M2 Max). Measurements capture the entire pipeline, from loading the NIfTI input to saving the extracted brain output and are based on the command-line implementation of MindGrab and the FreeSurfer version of SynthStrip. Benchmarking used the *pynvml* and *psutil* libraries for the NVIDIA GPU and the *time* utility on the Apple

M2 Max to capture memory usage and duration. Multiplicative factors (e.g., 2x) indicate MindGrab[⚡]'s efficiency gain over SynthStrip for each metric.

We benchmark MindGrab (through brainchop-cli), MindGrab⚡ (through brainchop-cli, with the --crop option), and SynthStrip (as mri_synthstrip of FreeSurfer) on an NVIDIA GeForce RTX 2080 GPU 11GB and Apple M2 Max GPU 64GB shared RAM, measuring RAM peak (GB), duration (s), as well as GPU memory (GB) for the NVIDIA GPU (see Figure 4). We ensure that pre- and post-processing operations guarantee results in the same space for both. On the Apple M2 Max, run duration and RAM usage were captured using the built-in *time* utility. For the NVIDIA GPU, we used the *pynvml* and *psutil* Python libraries to record the relevant metrics. Note that on the Apple M2 Max, GPU memory is shared with the main system RAM; therefore, this metric was not measured separately. Multiplicative factors are used to show efficiency gains and were determined by calculating the ratio of the medians of the two distributions being compared.

Panel (i) results consistently demonstrate MindGrab's lower RAM peak, shorter execution times, and reduced GPU memory usage compared to SynthStrip. On the NVIDIA GeForce RTX 2080, MindGrab achieves an approximately 4x lower RAM peak, a 2x speedup in run duration, and 2.3-3.1x lower GPU memory usage. The performance disparity is even more pronounced on the Apple M2 Max GPU, where MindGrab speedups range from 9.4-39.6x greater, and its RAM peak falls by approximately 25-33x compared to SynthStrip.

MindGrab⚡'s performance improvements over SynthStrip are even greater (Panel ii). On the NVIDIA GeForce RTX 2080, MindGrab⚡ achieves up to: 4x lower RAM peak, 2.6x speedup, and 4.9x lower GPU memory usage. On the Apple M2 Max GPU, speedups range from 13.2-42.4x greater, and peak RAM falls by 31.5-55.8x compared to SynthStrip.

### 3.4 Effects of Architectural Changes on Multimodal Performance

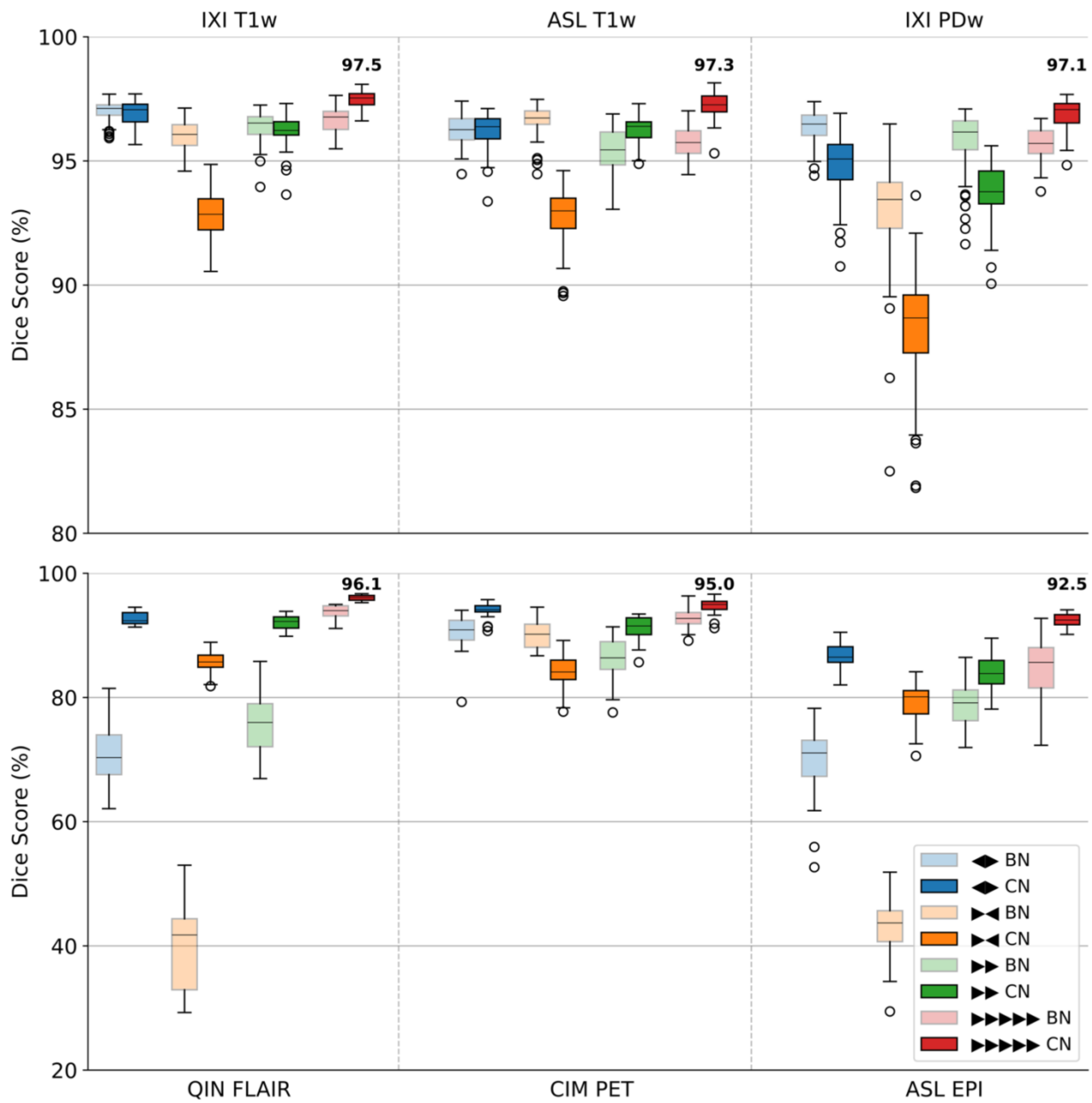

**Figure 5. Dice Score Distributions for Architectural Configuration and Normalization Comparison.** Dice score distributions for various architectural configurations for six datasets, sampled at the 0th, 20th, 40th, 60th, 80th, and 100th quantiles of MindGrab's performance (best to worst). Top row (left to right): 0%, 20%, 40%. Bottom row (left to right): 60%, 80%, 100%. Four configurations are shown (◀▶, ▶◀, ▶▶, ▶▶▶▶▶), with ▶ denoting a decreasing dilation sequence (16→8→4→2→1) and ◀ its reverse (1→2→4→8→16). Each configuration includes a BatchNorm (BN) and ChannelNorm (CN) variant. The BN variant is represented by a muted shade of its corresponding CN variant. The median Dice score for MindGrab (▶▶▶▶▶ CN) is displayed above its respective boxplot for each dataset. Note that the y-axis scale differs for the top and bottom figures.

We considered two strategies for combining the blurring and sharpening block sequences defined in section 2.1: autoencoder-like configurations (◀▶, ▶◀) and stacking identical blocks (▶▶, ▶▶▶▶▶). Figure 5 displays the performance distribution of these designs with BatchNorm and ChannelNorm variants across six multimodal datasets. ChannelNorm is implemented using Group Normalization (GroupNorm) with the number of groups set equal to the number of channels. This configuration standardizes each channel across the entire spatial extent of each 3D feature map. No learnable affine parameters are applied, making ChannelNorm a parameter-free normalization method. While BatchNorm is commonly used in segmentation literature, our observations indicate that ChannelNorm can lead to substantial performance improvements for MindGrab.

The autoencoder-like configurations exhibited dataset sensitivity. For instance, ◀▶ generally performed well for T1w datasets but struggled with other modalities such as ASL EPI and PDw, while ▶◀ showed greater variability across these datasets. Further refinement of these architectures was challenging due to key practical limitations: autoencoder-like dilation patterns do not lend themselves to effective stacking, and dilations beyond 16 offer no clear advantage (Section 2.1). These insights underscored the utility of the repeated stacking design.

While the ▶▶ model showed competitive performance with ◀▶ for the presented datasets, it did not consistently surpass it. However, the underlying design philosophy of repeated stacking—unreasonable in the autoencoder design—allowed us to extend the architecture further, leading to ▶▶▶▶▶ with parameter-free ChannelNorm, our proposed MindGrab model. MindGrab consistently achieves the highest Dice scores and exhibits the most robust performance among the evaluated configurations.

## 4. Discussion

Skull stripping remains a common yet challenging preprocessing step in neuroimaging pipelines. Our model, MindGrab, was designed with both accuracy and deployment feasibility as primary objectives. Its constant low memory footprint and low parameter count enable deployment as a lightweight command-line tool and in-browser model. We publicly release both versions through BrainChop under a permissive MIT license. The model (581KB) and full command-line package (<20MB with dependencies) are one-click installable on macOS, Linux, and Windows via brainchop-cli. The browser version requires no setup and can be accessed at brainchop.org. The model is identical across both platforms, ensuring that high accuracy is consistent regardless of deployment method. brainchop.org offers an intuitive user interface for single-instance analysis and quality control. The zero-footprint convenience of the browser version involves a performance trade-off, with runtimes that are longer (although still competitively fast) than the command-line tool and dependent on local hardware. The command-line tool is a better fit for large-scale data processing, leveraging its superior efficiency and support for batch operations to deliver significantly higher throughput than SynthStrip.

The performance advantage is particularly pronounced on non-NVIDIA hardware, such as Apple Silicon, demonstrating the tool's optimization for modern computing platforms beyond the specialized GPU ecosystem. This advantage is critical, as access to specialized GPUs remains limited in many research and clinical settings. MindGrab's lightweight architecture and high performance on common hardware, therefore, remove a primary bottleneck to the adoption of advanced AI tools. Ultimately, this democratizes access to state-of-the-art brain extraction, enabling its widespread and practical use in biomedical applications.

MindGrab's evaluation scores reveal its performance characteristics. MindGrab exhibits higher average precision than recall, indicating a conservative boundary delineation approach. In practice, this behavior suggests that MindGrab is more likely to minimally undersegment the brain compartment (as seen in the qualitative masks), prioritizing the accuracy of identified brain voxels. In contrast, MindGrab⚡ (cropped input) shifts this balance, leading to higher recall scores with a slight reduction in precision. This change implies a modified segmentation strategy more effective at capturing the whole brain, but with an increased propensity for including some non-brain tissue. This balance is aligned with SynthStrip's performance, which also generally favors recall over precision (overshoots) across most modalities. The availability of this adjustable trade-off between precision and recall through the default and cropped options of MindGrab's command-line implementation offers valuable flexibility for users, allowing them to select the mode that best fits their specific application and timing preference.

Qualitatively, MindGrab demonstrates highly effective and robust skull stripping for CT, PET, and MR scans, with its errors being uncommon and minimal across the evaluated samples. The primary difference between the model's prediction and the ground truth is typically over- and under-segmentation of the brain boundary. While these deviations can be genuine errors, they also frequently highlight the inherent ambiguity and subjectivity in defining a precise brain boundary, a challenge common to all skull stripping methods. To resolve undersegmentation differences and provide users with greater control, the brainchop-cli implementation of MindGrab includes a `--border` flag that allows users to increase the masks' boundary threshold in millimeters. This functionality ensures that users can achieve their desired segmentation while still benefiting from MindGrab's core advantages.

Our design prioritizes a spectral perspective over spatial receptive field size, which offers little practical guidance for architectures of this scale. The high-to-low dilation sequence (▶: 16→8→4→2→1) is key: its initial high-dilation layers are sensitive to high-frequency boundary details, while the subsequent low-dilation layers force this information into a progressively coarser, more robust representation. Cascading five such blocks (▶▶▶▶▶) creates a powerful information bottleneck, compelling the network to learn a highly efficient representation suitable for memory-constrained deployment. This design is complemented by the use of ChannelNorm (GroupNorm with groups equal to channels), a parameter-free standardization that empirically yielded greater robustness than standard batch normalization.

While MindGrab performed competitively against SynthStrip and outperformed ROBEX and BET, it has limitations with infant populations and segmenting images with artifacts that have high-contrast boundaries. Infant skull stripping challenges—including age-specific anatomical differences and lower tissue contrasts—leave pediatric performance unverified. Future research will aim to either train a dedicated pediatric version of MindGrab or develop a synthetic data generation protocol that more accurately captures the unique anatomical trends of infant populations. Nevertheless, its favorable performance efficiency trade-off makes MindGrab a practical tool for clinical and research applications.

**Funding Sources**

This work was supported by NIH 2R01EB006841 and in part by NSF 2112455. A. Fedorov was supported by the Nell Hodgson Woodruff School of Nursing at Emory University, C. Rorden by NIH awards P50-DC014664 and RF1-MH133701, and M. Hoffmann by NICHD grant R00 HD101553.

**Supplement**

**Table S1:** Mean Surface Distance to the Ground Truth ↓

| Modality | MindGrab | MindGrab⚡ | SynthStrip | ROBEX | BET |
|---|---|---|---|---|---|
| FSM T1w | 1.1 ± 0.1 | 1.1 ± 0.1 | 1.0 ± 0.1 | 1.7 ± 0.3 | 16.2 ± 4.8 |
| IXI T1w | 1.1 ± 0.1 | 1.2 ± 0.1 | 1.3 ± 0.2 | 1.5 ± 0.3 | 5.1 ± 3.0 |
| FSM qT1 | 1.1 ± 0.1 | 1.1 ± 0.1 | 1.0 ± 0.1 | 8.0 ± 5.8 | 15.9 ± 2.2 |
| ASL T1w | 1.2 ± 0.2 | 1.4 ± 0.2 | 1.2 ± 0.2 | 1.3 ± 0.4 | 4.3 ± 2.4 |
| FSM T2w | 1.3 ± 0.2 | 1.2 ± 0.2 | 1.0 ± 0.1 | 2.9 ± 0.9 | 3.5 ± 2.6 |
| IXI T2w | 1.2 ± 0.3 | 1.2 ± 0.3 | 1.4 ± 0.2 | 3.5 ± 1.2 | 2.9 ± 1.3 |
| IXI PDw | 1.3 ± 0.2 | 1.2 ± 0.3 | 1.4 ± 0.2 | 2.1 ± 0.5 | 2.2 ± 0.8 |
| IXI MRA | 1.2 ± 0.2 | 2.3 ± 0.5 | 1.0 ± 0.1 | 11.0 ± 4.0 | 4.5 ± 3.1 |
| FSM PDw | 1.5 ± 0.2 | 1.3 ± 0.2 | 1.1 ± 0.1 | 1.8 ± 0.4 | 5.8 ± 3.2 |
| QIN FLAIR | 1.5 ± 0.2 | 1.5 ± 0.1 | 1.5 ± 0.2 | 2.7 ± 1.8 | 1.7 ± 0.4 |
| CIM CT | 1.8 ± 1.0 | 2.0 ± 1.3 | 1.8 ± 0.4 | 11.5 ± 1.4 | 38.6 ± 4.7 |
| QIN T1w | 2.0 ± 0.6 | 2.1 ± 0.8 | 1.6 ± 0.4 | 2.7 ± 1.5 | 2.7 ± 1.2 |
| CIM PET | 2.0 ± 0.6 | 1.9 ± 0.5 | 1.9 ± 0.4 | 3.5 ± 1.5 | 4.5 ± 2.8 |
| IXI DWI | 2.4 ± 0.6 | 2.3 ± 0.6 | 1.7 ± 0.4 | 5.1 ± 2.8 | 2.3 ± 0.5 |
| QIN T2w | 2.4 ± 0.7 | 2.4 ± 0.7 | 1.8 ± 0.4 | 5.1 ± 3.1 | 3.9 ± 1.7 |
| ASL EPI | 2.6 ± 0.4 | 2.7 ± 0.4 | 1.6 ± 0.3 | 7.8 ± 3.1 | 1.9 ± 0.4 |

**Table S1**: Symmetric Mean Surface Distance (MSD ± standard deviation) between predicted and ground-truth brain masks for MindGrab, MindGrab⚡ (cropped input), SynthStrip, ROBEX, and BET across 16 diverse datasets. MindGrab, MindGrab⚡, and SynthStrip achieve low MSD values, indicating similarly accurate boundary delineation. ROBEX and BET show greater variability across modalities.

Table S1 reports the Symmetric Mean Surface Distance (MSD) to ground truth masks for each dataset, complementing the Dice scores and precision/recall relationships presented in Table 1 and Figure 1 of the main text. MSD is computed as the average of the minimum distance from each surface voxel in one mask to the surface voxels in the other mask, evaluated in both directions. While previously discussed metrics assess the volumetric overlap between predicted and ground truth masks, MSD quantifies how close the predicted brain boundary is to the ground truth boundary, highlighting boundary irregularities that might be averaged out in volumetric scores.

MindGrab, MindGrab⚡, and SynthStrip consistently achieve low MSD scores, indicating high boundary accuracy and surface conformity. SynthStrip has slightly lower values across several of the more challenging datasets, such as ASL EPI and IXI DWI, but both MindGrab and MindGrab⚡ remain within approximately one voxel of its performance. In contrast, ROBEX and BET scores exhibit substantial variability across datasets. While they perform competitively in isolated cases (e.g. BET on IXI DWI and ASL EPI; ROBEX on ASL T1w), their overall inconsistency highlights limitations in generalizing across imaging contrasts and modalities.

**Table S2:** 95th Percentile Symmetric Hausdorff Distance (HD95) ↓

| Modality | MindGrab | MindGrab⚡ | SynthStrip | ROBEX | BET |
|---|---|---|---|---|---|
| FSM T1w | 2.9 ± 0.4 | 2.7 ± 0.4 | 2.9 ± 0.4 | 4.6 ± 0.8 | 71.9 ± 16.1 |
| IXI T1w | 3.0 ± 0.7 | 3.0 ± 0.6 | 3.7 ± 0.6 | 4.7 ± 0.9 | 29.8 ± 17.3 |
| FSM qT1 | 3.0 ± 0.3 | 2.9 ± 0.3 | 2.9 ± 0.3 | 19.0 ± 10.9 | 75.7 ± 11.1 |
| ASL T1w | 3.0 ± 0.7 | 3.2 ± 0.7 | 3.4 ± 0.8 | 3.9 ± 2.2 | 21.8 ± 12.6 |
| FSM T2w | 3.2 ± 0.4 | 3.0 ± 0.4 | 2.8 ± 0.4 | 7.2 ± 3.5 | 17.5 ± 13.5 |
| IXI T2w | 3.2 ± 0.9 | 3.1 ± 0.8 | 3.8 ± 0.7 | 8.7 ± 3.1 | 9.2 ± 4.1 |
| IXI PDw | 3.3 ± 0.7 | 3.1 ± 0.8 | 3.7 ± 0.6 | 5.6 ± 1.9 | 8.2 ± 5.3 |
| IXI MRA | 3.3 ± 0.8 | 7.7 ± 2.4 | 2.6 ± 0.6 | 22.6 ± 4.4 | 13.8 ± 7.2 |
| FSM PDw | 3.6 ± 0.4 | 3.3 ± 0.4 | 3.0 ± 0.3 | 4.6 ± 1.6 | 35.8 ± 15.9 |
| QIN FLAIR | 4.3 ± 0.6 | 4.1 ± 0.5 | 4.4 ± 0.5 | 6.7 ± 4.2 | 5.1 ± 1.6 |
| CIM CT | 6.2 ± 7.0 | 8.1 ± 11.2 | 5.7 ± 1.5 | 20.5 ± 5.4 | 137.2 ± 15.5 |
| QIN T1w | 6.1 ± 2.4 | 7.1 ± 4.4 | 4.9 ± 1.5 | 8.3 ± 3.6 | 9.7 ± 4.9 |
| CIM PET | 5.1 ± 2.0 | 5.0 ± 1.9 | 5.3 ± 1.6 | 7.9 ± 2.4 | 15.0 ± 10.6 |
| IXI DWI | 6.7 ± 2.4 | 6.9 ± 2.3 | 4.8 ± 1.2 | 13.4 ± 5.9 | 7.1 ± 1.9 |
| QIN T2w | 8.3 ± 2.9 | 9.1± 3.6 | 5.7 ± 1.8 | 12.9 ± 5.5 | 15.2 ± 6.9 |
| ASL EPI | 7.7 ± 1.1 | 8.0 ± 1.3 | 4.6 ± 1.0 | 18.4 ± 4.0 | 4.9 ± 1.0 |

**Table S2**: 95th Percentile Symmetric Hausdorff Distance (HD95) across 16 imaging datasets spanning different modalities for the five evaluated brain extraction methods. MindGrab, MindGrab⚡, and SynthStrip achieve low HD95 scores, whereas ROBEX and BET achieve highly variable and large HD95 scores.

We report the 95th Percentile Symmetric Hausdorff Distance (HD95) in Table S2 to evaluate worst-case boundary errors and each model's robustness to localized segmentation failures. HD95 provides an upper bound on the geometric error encountered in 95% of all boundary points, quantifying the severity of the largest segmentation error while mitigating the influence of isolated outliers.

MindGrab, MindGrab⚡, and SynthStrip consistently achieve lower HD95 scores. Performance between MindGrab and MindGrab⚡ is generally comparable, with each showing slight advantages on different datasets consistent with trends observed in other evaluation metrics. Conversely, ROBEX and BET exhibit substantial variability and large HD95 scores across most modalities and datasets, reflecting inconsistent performance and reduced generalizability.